\documentclass[]{aa}
\usepackage[varg]{txfonts}
\usepackage{graphicx}
\usepackage{textcomp}
\usepackage{color}
\usepackage{bbold}
\usepackage{breqn}
\usepackage{amsmath}
\usepackage{pstricks} 
\usepackage{bm} 
\usepackage{relsize}

\usepackage[colorlinks=true,urlcolor=blue,citecolor=blue,linkcolor=blue,breaklinks=true]{hyperref}
\usepackage{twoopt}

\def\xlinkspace#1 #2{%
 \ifx\relax#2%
 \xlinkdash#1-\relax
 \else
 \xlinkdash#1 -\relax
 \expandafter\xlinkspace\expandafter#2%
 \fi}

\def\xlinkdash#1-#2{%
 \ifx\relax#2%
 \tmp{#1}%
 \else
 \tmp{#1-}%
 \expandafter\xlinkdash\expandafter#2%
 \fi}

\newcommand\CC{C\nolinebreak[4]\hspace{-.05em}\raisebox{.4ex}{\relsize{-3}{\textbf{++}}}}

\bibpunct{(}{)}{;}{a}{}{,} 

\makeatletter

 \newcommandtwoopt{\citeads}[3][][]{%
   \nonstopmode
   \href{http://adsabs.harvard.edu/abs/#3}%
        {\def\hyper@linkstart##1##2{}%
         \let\hyper@linkend\@empty\citealp[#1][#2]{#3}}
   \biblink{#3}{\href{http://adsabs.harvard.edu/abs/#3}{ADS}}%
   \errorstopmode}            
   
 \newcommandtwoopt{\citepads}[3][][]{%
   \nonstopmode
   \href{http://adsabs.harvard.edu/abs/#3}%
        {\def\hyper@linkstart##1##2{}%
         \let\hyper@linkend\@empty\citep[#1][#2]{#3}}
   \biblink{#3}{\href{http://adsabs.harvard.edu/abs/#3}{ADS}}
   \errorstopmode}            
   
 \newcommandtwoopt{\citetads}[3][][]{%
   \nonstopmode
   \href{http://adsabs.harvard.edu/abs/#3}
        {\def\hyper@linkstart##1##2{}%
         \let\hyper@linkend\@empty\citet[#1][#2]{#3}}
   \biblink{#3}{\href{http://adsabs.harvard.edu/abs/#3}{ADS}}%
   \errorstopmode}            
   
 \newcommandtwoopt{\citeyearads}[3][][]{%
   \nonstopmode
   \href{http://adsabs.harvard.edu/abs/#3}%
        {\def\hyper@linkstart##1##2{}%
         \let\hyper@linkend\@empty\citeyear[#1][#2]{#3}}
   \biblink{#3}{\href{http://adsabs.harvard.edu/abs/#3}{ADS}}%
   \errorstopmode}            
\makeatother

\makeatletter
\newcommand{\bibnote}[2]{\@namedef{#1note}{#2}}
\newcommand{\biblink}[2]{\@namedef{#1link}{#2}}
\makeatother

\newcommand{\be}{\begin{equation}}
\newcommand{\ee}{\end{equation}}

\newcommand{\bea}{\begin{eqnarray}}
\newcommand{\eea}{\end{eqnarray}}

\begin{document}

\title{{STiC} -- A multiatom non-LTE PRD inversion code for full-Stokes solar observations}

\subtitle{ }

\author{
  J. de la Cruz Rodr\'{i}guez\inst{1} \and
    J. Leenaarts\inst{1} \and
    S. Danilovic\inst{1} \and
  H. Uitenbroek\inst{2}  %
}

\offprints{J. de la Cruz Rodr\'iguez \email{jaime@astro.su.se}}

\institute{
Institute for Solar Physics, Dept. of Astronomy, Stockholm University, AlbaNova University Centre, SE-106 91 Stockholm, Sweden
\and
National Solar Observatory, 3665 Discovery Drive, Boulder, CO 80303, USA
}
\titlerunning{STiC}
\authorrunning{de la Cruz Rodr\'iguez et al. }

\date{Received; Accepted }

\abstract 
{The inference of the underlying state of the plasma in the solar chromosphere remains extremely challenging because of the nonlocal
  character of the observed radiation and plasma conditions in this layer. Inversion methods allow us to derive a model atmosphere that can
  reproduce the observed spectra by undertaking several physical assumptions.
  
  The most advanced approaches involve a depth-stratified model atmosphere described by
temperature, line-of-sight velocity, turbulent velocity, the three components of the magntic field vector, 
and gas and electron pressure. The parameters of the radiative transfer equation are computed from 
a solid ground of physical principles. In order to apply these techniques to spectral lines that sample
the chromosphere, nonlocal thermodynamical equilibrium effects must be included in the calculations.

We developed a new inversion code STiC (STockholm inversion Code) to study spectral lines that sample the upper 
chromosphere. The code is based on the RH forward synthesis code, which we modified to make the inversions faster and 
more stable. For the first time, STiC facilitates the processing of lines from multiple atoms in non-LTE, also including partial redistribution 
effects {(PRD)} in angle and frequency of scattered photons.
Furthermore, we include a regularization strategy that allows for  model atmospheres
with a complex depth stratification, without introducing artifacts in the reconstructed physical parameters, which are usually manifested in the form of oscillatory behavior. This approach takes steps toward a node-less inversion, in which the value of the physical parameters at each grid point can be considered a free parameter.

In this paper we discuss the implementation of the aforementioned techniques, the description of the model atmosphere,
 and the optimizations that we applied to the code. We carry out some numerical experiments to show the performance 
 of the code and the regularization techniques that we implemented. We made
 STiC publicly available to the community.
}

    \maketitle

\section{Introduction} \label{sec:intro}
The new generation of 4 m telescopes (DKIST, EST) aims at studying the chromosphere and its coupling to the underlying photosphere with unprecedented spatial resolution and signal-to-noise ratio. In the photosphere the local thermodynamical equilibrium approximation (LTE) can be adopted to model the observations in most spectral lines, but in the chromosphere collisional rates are in comparison very low, making this assumption generally not valid. Therefore the translation of the observed intensities to the underlying physical parameters of the plasma remains very challenging and the nonlocal character of the problem must be taken into account. From an observational perspective one of the most successful approaches to do so has been {spectropolarimetric inversions} (hereafter inversions; see reviews by \citeads{2016LRSP...13....4D} and \citeads{2017SSRv..210..109D}).

Inversion codes allow the reconstruction of physical parameters from spectropolarimetric observations by assuming a model. Traditionally, the assumed model can be described by parameters of the radiative transfer equation directly (e.g., Milne-Eddington or constant slab model), or with thermodynamical variables from which the parameters of the radiative transfer equation can be calculated. {The former does not allow the inclusion of a depth-varying line-of-sight velocity of magnetic field vector. Depth-varying LTE inversions} were introduced by \citetads{1992ApJ...398..375R} under the assumption of LTE in the Stokes Inversion based on Reponse functions code (SIR). Under these conditions the atom populations are strictly set by the local conditions of the atmospheric plasma and the computation of the emerging intensity can be computed {by calculating a formal solution of the polarized radiative transfer equation}. These ideas were also used in the {Stokes-Profiles-INversion-ORoutines} (SPINOR{;} \citeads{2000A&A...358.1109F}). {Perhaps the most popular photospheric lines used in {Milne-Eddington} (ME) and LTE inversions are \ion{Fe}{i} lines such as the $525$~nm dublet,  $617$~nm, $630$~nm dublet, and $1565$~nm dublet (some examples of recent studies are \citeads{2013A&A...553A..63S}; \citeads{2014A&A...569A.105J}; \citeads{2015A&A...576A..27B}; \citeads{2016A&A...596A...5M}; \citeads{2016ApJ...832..170P}; \citeads{2016A&A...593A..93D}; \citeads{2017ApJS..229....3C}; \citeads{2017ApJS..229....5D}; \citeads{2017A&A...601L...8B}; \citeads{2018A&A...616A..46P}).}
\begin{figure*}
\centering
\includegraphics[width=\textwidth]{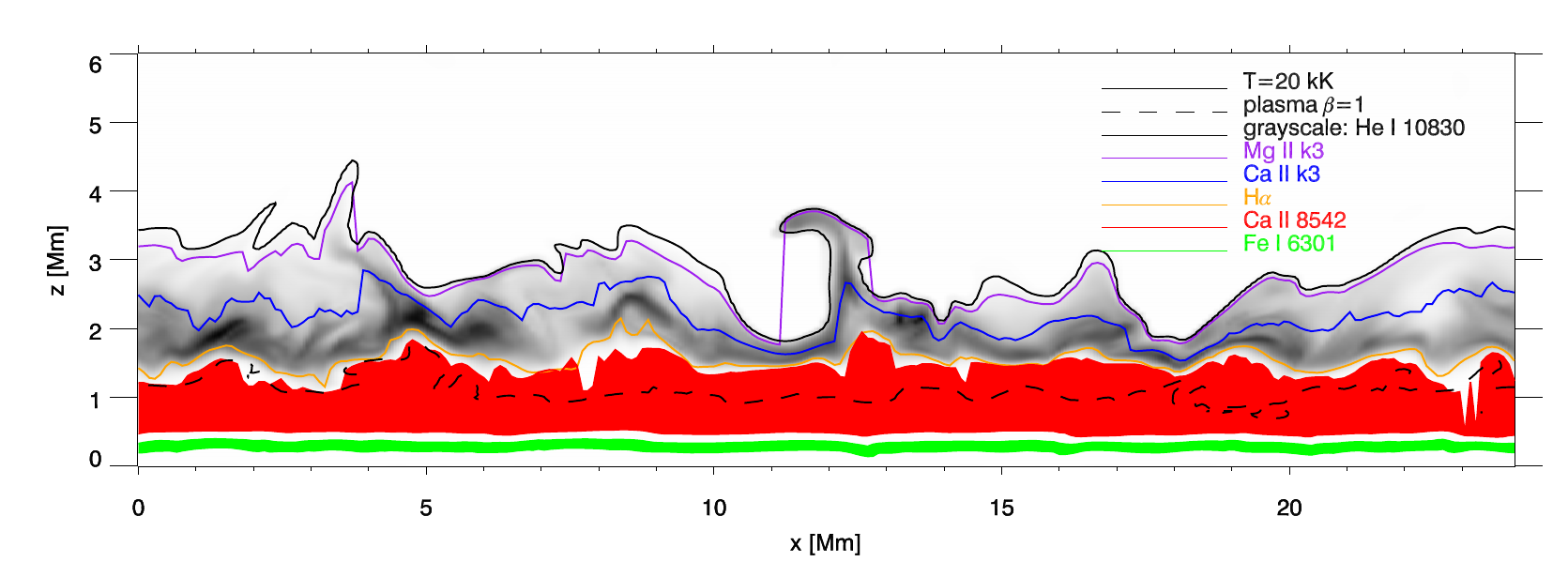}
\caption{Vertical slice from a Bifrost radiation-MHD simulation, indicating the approximate formation height of different diagnostics. The solid lines indicate the $\tau=1$ layer at the core of the \ion{Mg}{ii}~k line (purple),  \ion{Ca}{ii}~K line (navy) and  H$\alpha$ line (orange). The black solid line indicates the location where $T=20$~kK and the dashed black line the plasma $\beta = 1$ layer. The gray shades illustrate the \ion{He}{i}~10830 line opacity. We indicated the entire formation range of the \ion{Fe}{i}~6301 line in green and the \ion{Ca}{ii}~8542 line in red.}
\label{fig:layers}
\end{figure*}

The first attempts to perform such calculations under non-LTE conditions, where the rate equations are not dominated by collisional terms, were carried out by \citetads{2000ApJ...530..977S} in the \ion{Ca}{ii} infrared triplet lines ($\lambda 8498, \lambda 8542, \lambda 8662$). Currently, the {Non-LTE Inversion Code based on the Lorien Engine (NICOLE)} allows us to perform such inversions assuming Zeeman-induced polarization (\citeads{2015A&A...577A...7S}). Only recently a new non-LTE inversion code has been developed that includes for the first time non-LTE analytical response functions (\citeads{2018A&A...617A..24M}).

The \ion{Ca}{ii} infrared triplet lines sample the lower chromosphere (see Fig.~\ref{fig:layers}) and they can be modeled assuming a complete redistribution of scattered photons (\citeads{1989A&A...213..360U}) and statistical equilibrium (\citeads{2011A&A...528A...1W}). {Comparatively, the \ion{Na}{i}~D lines and \ion{Mg}{i}~$517$~nm line have larger sensitivity to magnetic fields in the upper photosphere/lower chromosphere, but they form deeper than the \ion{Ca}{ii}~IR triplet lines (\citeads{2010ApJ...709.1362L}; \citeads{2011A&A...531A..17R}; \citeads{2018MNRAS.481.5675Q})}.

Unfortunately, the selection of lines that are sensitive to the upper chromosphere is not large; these are \ion{Mg}{ii}~h\& k, \ion{Ca}{ii}~H\&K, the Lyman alpha line, and the \ion{He}{i}~D$_3$ and 10830 lines. With the exception of the \ion{He}{i} lines, all the aforementioned diagnostics are strong resonance lines that are affected by partial redistribution effects of scattered photons, and these effects must be included in forward synthesis calculations. However, there are good reasons to attempt to include some of these lines simultaneously in one inversion, which has not been possible until now. These reasons are as follows: 
\begin{enumerate}
\item By including information of the upper chromosphere, we can attempt to discriminate between physical processes that do not leave a clear imprint in the lower chromosphere (e.g., \citeads{2016ApJ...827..101K}){, or attempt to constrain opacity effects induced by the combined action of gas flows and temperature fluctuations (\citeads{1984mrt..book..173S}; \citeads{2015ApJ...810..145D}; \citeads{2017ApJ...845..102H})}.
\item These upper chromosphere lines are also sensitive to the middle and lower chromosphere, providing valuable redundant information to constrain physical parameters in those layers (see Fig.~\ref{fig:layers}).
\item If spectral lines from different atomic species can be processed simultaneously, some of the degeneracies that can arise between temperature (opacity broadening) and microturbulence (e.g., \citeads{1974SoPh...39...49S}; \citeads{2015ApJ...809L..30C}) can be ameliorated because the thermal term present in the Doppler broadening of the line is divided by a different mass, whereas the turbulent velocity term is the same in all cases.
\end{enumerate}

In this paper we present the STockholm inversion COde (STiC), which allows, for the first time, the processing of spectral lines from multiple atoms simultaneously, including partial redistribution effects (PRD) of scattered photons in angle and frequency. In the present paper, we discuss how the code operates and we present a new regularizing Levenberg-Marquardt algorithm (LM), which improves the convergence of the inversion process and allows for more degrees of freedom without loosing fidelity and unicity of the solution. Early versions of STiC have {already} been used in previous studies (\citeads{2016ApJ...830L..30D}; \citeads{2018A&A...612A..28L}; \citeads{2018ApJ...857...48G}; \citeads{2018arXiv180606682D}).

\section{Spectral synthesis module}\label{sec:rh}
The STiC code started as a modular LTE inversion code. {In order to add non-LTE radiative transfer capabilities}, we  modified the 2014 version of RH (\citeads{2001ApJ...557..389U}) to create a synthesis module that can be called efficiently from the main inversion code. The RH code can solve the non-LTE problem in multiple atoms at the same time, and it includes PRD effects in strong resonance lines. 

The code STiC operates on a column-by-column basis, assuming plane-parallel geometry to solve the statistical equilibrium equations, usually referred to as the 1.5D approximation. Once the atom population densities are known for all atoms, a final formal solution is computed at the heliocentric angle of the 
observations. Unfortunately, horizontal radiative transfer cannot be included in these kinds of calculations as the computation of the derivatives of the intensity vector
with respect to each physical parameter would be prohibitive, among other challenges such as radiative coupling that would be present among the different pixels.

Scattering is expected to be more dominant when collisional rates are low, and the latter are particularly low in the upper chromosphere. A number of recent studies have emphasized the importance of 3D radiative transfer when modeling strong scattering lines (e.g., \citeads{2009ApJ...694L.128L}; \citeads{2017A&A...597A..46S}; \citeads{2018A&A...611A..62B}), especially for simulating observations toward the solar limb.
However, those studies have made use of radiation 3D magnetohydrodynamics (MHD) simulations representative of quiet-Sun situations. In fact, the chromospheric gas density in those MHD simuilations seems to be lower than what observations indicate, even in the quiet-Sun (see details in \citeads{2016A&A...585A...4C}). The main target for our inversions are active regions, where the magnetic field is stronger and the ionization degree is higher than in the quiet-Sun. Furthermore, in active regions the transition region is usually pushed to higher mass densities and the local temperature is larger than in the quiet-Sun (\citeads{2015ApJ...809L..30C}). All these effects would arguably lead to larger collisional rates in the chromosphere than the situation represented in those MHD simulations. For all these reasons we are compelled to speculate that the aforementioned studies that analyze 3D effects are representative of a worse case scenario when active regions are the main observational target.

\subsection{Changes and optimizations to RH}
This version of RH includes the fast PRD angle approximation proposed by \citetads{2012A&A...543A.109L}, but we optimized the original algorithm with the following changes:
\begin{itemize}
\item The algorithm originally implemented in RH by \citetads{2012A&A...543A.109L} computed the mean intensity in the comoving frame of the grid cell for all wavelengths, but that quantity is only used in calculations related to PRD or {cross-redistribution lines (XRD)}. We changed the structure of the algorithm to ensure that these operations are only performed and stored for frequencies associated with PRD/XRD lines, rather than for the entire emerging spectrum.
\item We rearranged and restricted the extent of the loops where the interpolation coefficients are computed, saving between a factor two and three in the execution time of these calculations.
\end{itemize}
With this new implementation of the algorithm, the amount of time spent in the precomputation of interpolation coefficients is negligible in most applications. 

RH can compute the {van der Waals} damping parameter (from collisions with neutral hydrogen) using the recent formalism of \citetads{2000A&AS..142..467B}. Inside RH, this is done by interpolating the corresponding coefficients for the line under consideration from a table that is only valid for neutral species. For lines from ionized species these tables cannot be used. However, these coefficients have been computed for strong chromospheric lines like \ion{Ca}{ii}~H\&K, the \ion{Ca}{ii}~IR triplet lines, and \ion{Mg}{ii}~h\&k (e.g., \citeads{1998MNRAS.300..863B}) and they allow for a more accurate estimate of the damping wings. Therefore, we slightly modified the input atom format to allow for the feeding of those coefficients manually if needed.

Finally, since we need to compute response functions by finite differences during the inversion,
we allow the code to store departure coefficients from LTE that can be used to initialize the atom populations during the inversions, a trick that was already introduced in the NICOLE code. This simple change allows the response functions at each node to be computed with very few iterations.

\subsection{Formal solution of the radiative transfer equation}
\begin{figure}
\centering
\includegraphics[width=\columnwidth]{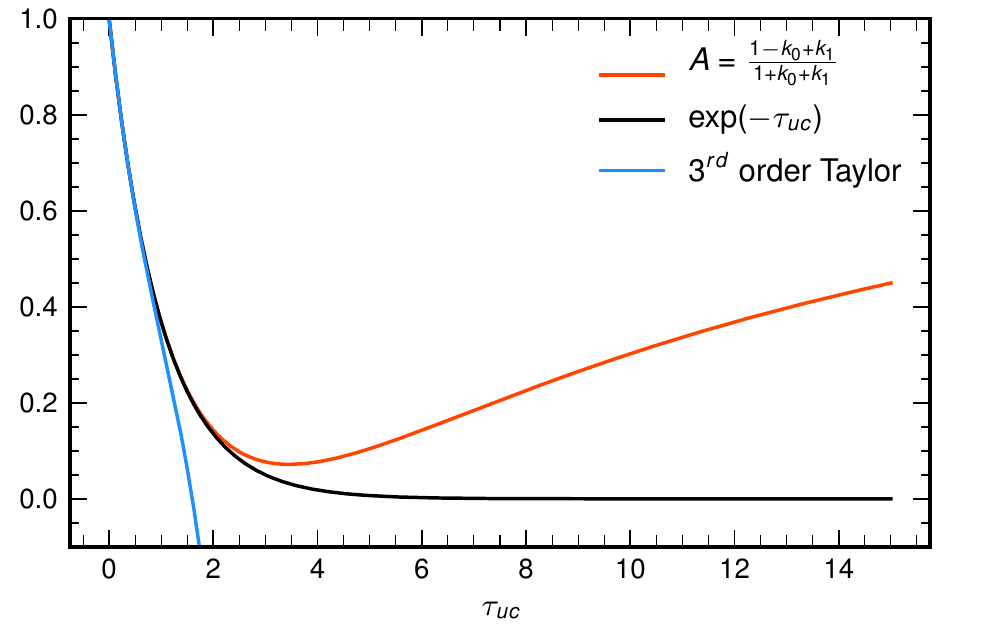}
\caption{Comparison of the integration coefficient $A$ from LBR's method (red) with an exponential (black). $A$ is a good approximation to the exponential for small optical thickness, but it greatly deviates close to the optically thick regime.}
\label{fig:coeff}
\end{figure}

We included cubic DELO-Bezier formal solvers for polarized and unpolarized radiation (\citeads{2013ApJ...764...33D}), which allow us to accurately solve the radiative transfer equation in coarse depth grids. Given our choices in the definition of the Bezier interpolant control points used in the formal solver, the latter is exactly equivalent to a Hermite method (\citeads{2003ASPC..288....3A}; \citeads{2013A&A...549A.126I}). The accuracy of these methods has been recently analyzed in great detail for the polarized case by \citetads{2017ApJ...845..104J} and \citetads{2018ApJ...857...91J}.

We found an unfortunate mistake in the transcription of the cubic Bezier integration coefficients reported by \citetads{2013ApJ...764...33D}. The monochromatic unpolarized cubic Bezier integration scheme is given by
\begin{equation}
        I_c = I_u \mathrm{e}^{-\tau_{uc}} + S_u\alpha +  S_c\beta + C_u\gamma + C_c\varphi, \label{eq:bezier}
\end{equation}
where the subindex $u$ indicates quantities located in the upwind point, where both the intensity ($I_u$), the source function ($S_u$) and the control point ($C_u= S_u + S'_u\tau_{uc}/3$) are known. The subindex $c$ indicates quantities located in the central point, where only the source function $S_c$ and the control point ($C_c = S_c - S'_c\tau_{uc}/3$) are known, but not the intensity ($I_c$) that we want to compute.
The corrected interpolation coefficients are
\begin{eqnarray}
\alpha &=& -\frac{\mathrm{e}^{-\tau_{uc}}(6-6\tau_{uc}+3\tau_{uc}^2+\tau_{uc}^3) - 6}{\tau_{uc}^3},\nonumber\\
\beta  &=& \frac{-6+\tau_{uc}(6+\tau_{uc}(\tau_{uc}-3)) + 6\mathrm{e}^{-\tau_{uc}}}{\tau_{uc}^3},\label{eq:bezC}\\
\gamma &=& 3\cdot\frac{2\tau_{uc} -6 +\mathrm{e}^{-\tau_{uc}}(6+\tau_{uc}(\tau_{uc}+4))}{\tau_{uc}^3},\nonumber\\
\varphi &=& 3\cdot\frac{-2\mathrm{e}^{-\tau_{uc}}(\tau_{uc}+3) + 6 +\tau_{uc}(\tau_{uc}-4)}{\tau_{uc}^3}\nonumber.
\end{eqnarray}

In their analysis, \citetads{2017ApJ...845..104J} also considered another {Hermitian} method, originally introduced by \citetads{1998ApJ...506..805B} (LBR hereafter), which seems to perform extremely well in their tests. We were encouraged by those results to try to implement polarized and unpolarized version of these solvers. The latter does not make use of the analytical formal solution of the transfer equation which, for unpolarized light, is
\begin{equation}
I_c = I_u \mathrm{e}^{-\delta\tau_\nu} + \int_0^{\tau_{uc}}S(t)\mathrm{e}^{-(\tau_{uc}-t)}dt,\label{eq:formal}
\end{equation}
where $I_c$ is the intensity at the {central point}, $I_u$ is the already computed upwind intensity, $S$ is the source function, and  $\delta\tau_{uc}$ is the optical thickness of the medium between the upwind and central points. Normally, the integral of the source function in Eq.~(\ref{eq:formal}) is solved analytically by approximating the source function with a given depth-dependence: linear, parabolic, Bezier, Hermite. So these interpolants are only used to describe the source function.

In LBR's method, the intensity vector at the central point of the interval is computed with a polynomial expansion around the upwind point. After some algebra, and using the transfer equation to provide the first and second derivatives of the intensity, they obtained a solution with Hermitian form.

A priori this method is very fast to compute and it does not require the computation of exponentials and vanishing small quantities. The latter may be a great asset when computing the mean intensities that are used to solve the non-LTE problem. But it does so by loosing some of the insights provided by the analytical form of the transfer equation in Eq.~(\ref{eq:formal}). The reason is that in this case, the polynomial expansion of the intensity must account for the exponential term in the optically thin regime, but it must also properly describe the behavior of the local contributions in optically thick cases.

If we apply the principles described in \citetads{1998ApJ...506..805B} to the unpolarized case, we recover the usual integration scheme that depends on the ensuing intensity, the source function values, and its derivatives as follows:
\begin{equation}
  I_c = I_uA + \bar{\alpha} S_u + \bar{\beta} S_c + \bar{\gamma} (S'_u-S'_c)\label{eq:lbr}
,\end{equation}
where the integration coefficients are similarly defined in terms of $\tau_{uc}$. Defining $ k_0 = \tau_{uc} / 2$ and $k_1 = (\tau_{uc})^2 / 12$ we can express those coefficient as
\begin{eqnarray*}
  A &=& \frac{1-k_0+k_1}{1+k_0+k_1},\\
  \bar{\alpha} &=& \frac{k_0-k_1}{1+k_0+k_1},\\
  \bar{\beta} &=& \frac{k_0+k_1}{1+k_0+k_1},\\
  \bar{\gamma} &=& \frac{k_1}{1+k_0+k_1}.
\end{eqnarray*}
Comparing Eq.~(\ref{eq:lbr}) to Eq.~(\ref{eq:formal}), we would expect the coefficient $A$ to behave like $\mathrm{e}^{-\tau_{uc}}$. Fig.~\ref{fig:coeff} shows the comparison of these two quantities as a function of $\tau_{uc}$. In fact, coefficient $A$ coincides with a second order Pad\'e approximant \citep{pade1892representation} of the exponential function around $\tau_{uc}\approx 0$, which is usually considered to be a more accurate and somewhat better behaved approximant than the Taylor expansion (see Fig~\ref{fig:coeff}), but it is still a polynomial approximation. For small optical thickness, coefficient $A$ accurately follows the exponential, but from $\tau_{uc}>0.1$ the error is numerically noticeable and it starts to deviate from the exponential behavior and to even increase from $\tau_{uc}\gtrapprox3.4$, giving unrealistically high weight to the incoming radiation even when the medium is optically thick (where that part should be attenuated by the exponential). 

If the scheme presented in Eq.~(\ref{eq:lbr}) is used to solve the non-LTE problem with strong scattering lines, the latter does not converge because the scheme is not accurate in the region of interest for these lines ($\tau_{uc}\lessapprox 10$). A similar conclusion may be applied to the polarized case, because it is also based on a polynomial expansion of the intensity vector around the upwind point, and that approximation is only supposed to work for small excursions from the point where the intensity is approximated.

Pad\'e approximants can be more precise than a Taylor expansion of the same order for larger excursions from the origin point, as suggested by Fig.~\ref{fig:coeff}. {In Appendix~\ref{sec:pade} we provide Pad\'e polynomial approximations of the cubic Bezier interpolation coefficients, which are valid for small optical tickness regimes, where high-order schemes usually suffer from numerical errors.}

\section{Inversion engine} \label{sec:clm}
\subsection{Equation of state}\label{sec:model}
The STiC code works with depth-stratified atmospheres including the stratification of temperature, gas pressure, electron density, line-of-sight velocity, microturbulence, and the three components of the magnetic field vector as {functions} of column mass, optical depth or height. The magnetic field vector is decomposed in the longitudinal component ($B_\parallel$), strength of the transverse component ($|B_\bot|$) and azimuth of the transverse component ($B_\chi$).
\begin{figure}
\centering
\includegraphics[width=\columnwidth]{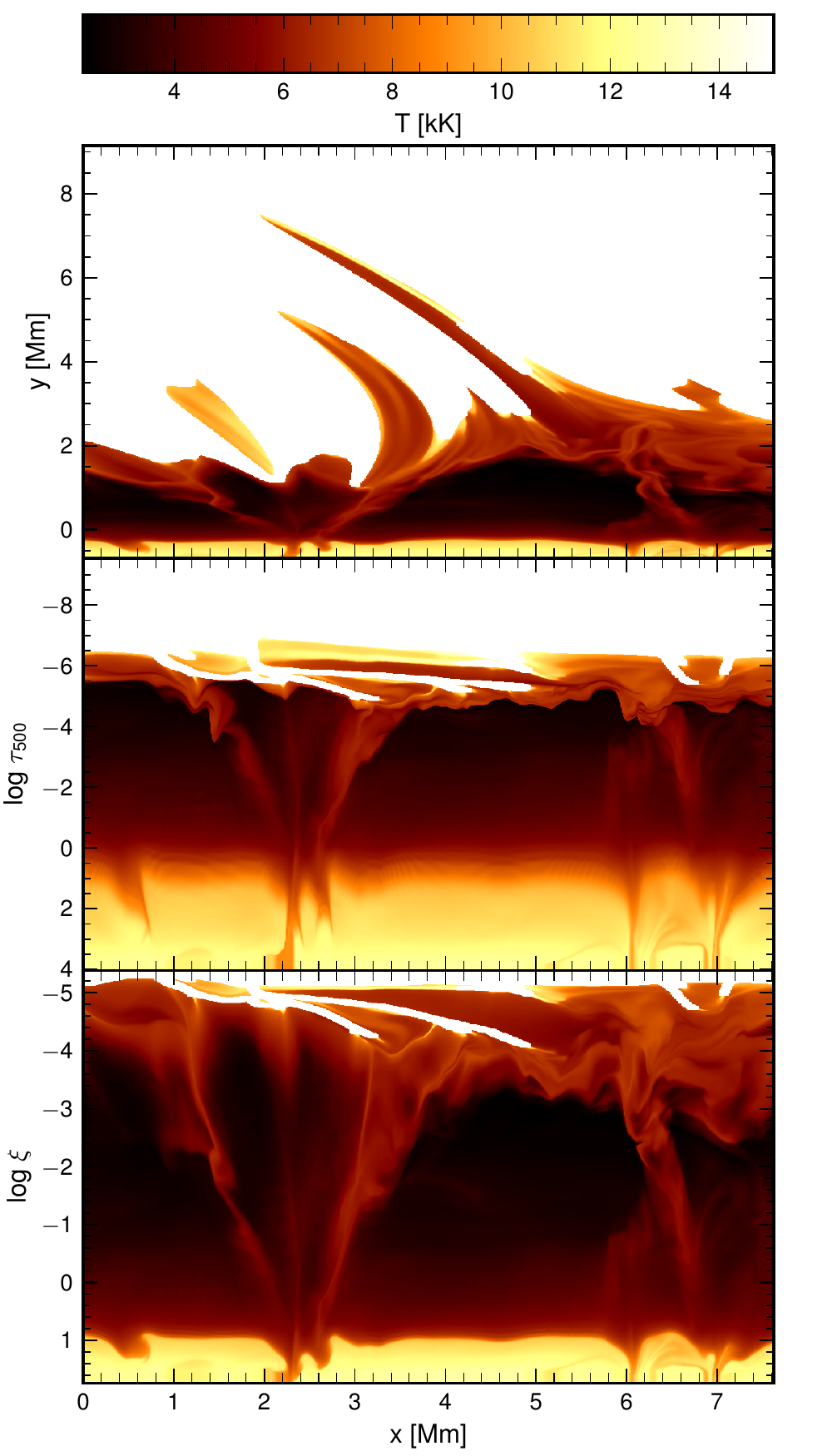}
\caption{Vertical slice of a temperature snapshot from a publicly available 2.5D rMHD simulation (\citeads{2017Sci...356.1269M}). We adopted the original geometrical depth scale (top), an optical-depth scale (middle), and a column mass depth scale (bottom) to represent the same temperature slice. }
\label{fig:bif}
\end{figure}

The inversion engine parameterizes the stratification of physical quantities as a function of optical depth or column mass. If the gas pressure and electron densities are not known, they can be derived assuming a gas pressure value at the upper boundary of the atmosphere and integrating the hydrostatic equilibrium equation. During the inversion, the gas pressure stratification and electron densities are derived assuming hydrostatic equilibrium, but in pure synthesis mode they can be provided externally. 
Similar to other inversion codes (e.g., SIR, NICOLE), we solve the equation of hydrostatic equilibrium to derive the gas pressure scale for any {guessed model} atmosphere (\citeads{1970stat.book.....M}) as follows:
\begin{equation}
        \frac{\partial p}{\partial\tau_{\lambda}} = -\frac{g}{\kappa_{\lambda} + \sigma_{\lambda}},\label{eq:hydro}
\end{equation}
where $p$ is gas pressure, $\tau_\nu$ is the optical depth, $g$ is the value of gravity,  $\kappa_{\nu}$ is the mass absorption, and $\sigma_{\lambda}$ is the scattering coefficient for continuum opacity. The subindex $\nu$ {refers} to a reference wavelength, typically 500~nm. The parameters $\kappa_\lambda$ and $\sigma_\lambda$ are computed assuming a LTE equation of state and background continuum opacity. This equation of state was adopted from the SME: evolution code (\citeads{2017A&A...597A..16P}). Eq.~(\ref{eq:hydro}) can be integrated numerically assuming linear dependence of $\beta_\lambda = \kappa_\lambda + \sigma_\lambda$ between consecutive grid cells. To simplify the notation we assume a discrete grid of $k=1,..,n_{dep}$ values, where index 1 represents the upper boundary of the atmosphere and $\lambda=500$~nm. Since $\beta_k$ depends on the value of $p_k$, a few iterations are needed to get the values of $\beta_k$ and $p_k$ to be consistent. In that case, the solution is written as\begin{equation}
        p_{k} = p_{k-1} + 
        \begin{cases}
        \frac{g(\tau_{k} - \tau_{k-1})}{\beta_{k-1}} & iter = 1\\
        \frac{g(\tau_{k} - \tau_{k-1})}{\beta_k - \beta_{k-1}} \log \left ( \frac{\beta_k}{\beta_{k-1}} \right) & iter > 1
        \end{cases}     
.\end{equation}
If, on the contrary we decide to perform the inversion using column mass ($\xi$) as a depth variable, the hydrostatic equilibrium equations simplify greatly, no opacity calculations are involved, and no iterations are needed (see, e.g., \citeads{2014tsa..book.....H}), i.e.,
\begin{equation}
p_{k} =g\xi.
\end{equation}
Working in any of these depth scales is somewhat equivalent to using a Lagrangian frame as the physical quantities follow density (not strictly in the case of optical depth, but very closely related). Working in column mass naturally sets the boundary condition at the top of the atmosphere $p_{k=0} = g\xi$. The main difference between working in column mass or optical depth is that the former stretches the chromosphere a bit more and greatly compresses the transition region, whereas the latter comparatively allows us to better resolve the transition region and photosphere (see Fig.~\ref{fig:bif}). This figure gives some insight about how spicules would be squeezed in an optical depth or in a column-mass scale. They would appear as a localized bump in temperature because these scales are not sensitive to the coronal plasma that surrounds these cold material protrusions.

The electron pressure is tightly related to the local gas pressure and temperature. Once the gas pressure is known, we initialize the electron pressure under the assumption of LTE. Hydrogen is the main electron donor in the chromosphere. In principle, if hydrogen is included as an active non-LTE species in RH, the electron density can be iterated internally in RH to make it consistent with the non-LTE hydrogen ionization (in statistical equilibrium). We modified RH to allow us to solve the statistical equilibrium equations along with charge conservation (e.g., \citeads{2007A&A...473..625L}) for the hydrogen atom. With the modified equations, Newton-Raphson iterations are needed within each {multi-level accelerated lambda} iteration {(MALI)} to make the electron density and hydrogen ionization  consistent with each other.

The penalty of including hydrogen as an active species is large. In that case, the whole process becomes up to eight times slower compared to the LTE case. We only encourage the use of this setup for selected pixels, and we always recommend starting from a relatively converged atmosphere from an inversion with hydrogen in LTE.

\subsection{Atmospheric parameterization}\label{sec:nodes}
\citetads{1992ApJ...398..375R} introduced depth-stratified inversions based on {nodes}. These nodes represent the free parameters of our model. The inversion modifies the values of the nodes and generates a new fully stratified atmosphere that can be used to solve the (polarized) radiative transfer equation. 

Radiative transfer calculations require a relatively dense grid of depth points to solve accurately the radiative transfer equation numerically. \citetads{1992ApJ...398..375R} showed that the inversion cannot be performed in such a fine grid because the observables would not constrain all these degrees of freedom. Therefore, they introduced a coarser grid {(the nodes)} for each physical parameter. They used piece-wise segments or splines to connect the nodes in the finer grid.

{However, there is a fundamental difference in our implementation compared with the node approach introduced by \citetads{1992ApJ...398..375R}. In the latter case, the nodes were used to describe corrections that are interpolated into the fine grid and added to an input model atmosphere. In our case, the nodes represent the actual value of the model atmosphere, that is directly interpolated into the fine depth grid. {The \citetads{1992ApJ...398..375R} approach} allows inversions to be performed with a relatively low number of nodes if the {structuring} of the input atmosphere is close to {that of} the final result. However, if the input model has a complex structure that does not correspond to the profiles that we are inverting, this approach struggles to remove such structuring, even with many nodes. {Our} approach is somewhat safer because the depth complexity is directly set by the number of nodes. We implemented both approaches in STiC, however we recommend using the scheme in which the nodes represent the actual value of the atmospheric parameter. The SPINOR2D, SPARSE and SNAPI codes also make use of this node scheme (\citeads{2012A&A...548A...5V}; \citeads{2015A&A...577A.140A}; \citeads{2018A&A...617A..24M}), which also simplifies the implementation of spatial coupling and regularization that operate directly on the model parameters (see \S\ref{sec:LM})}.

\begin{figure}
\centering
\includegraphics[width=\columnwidth]{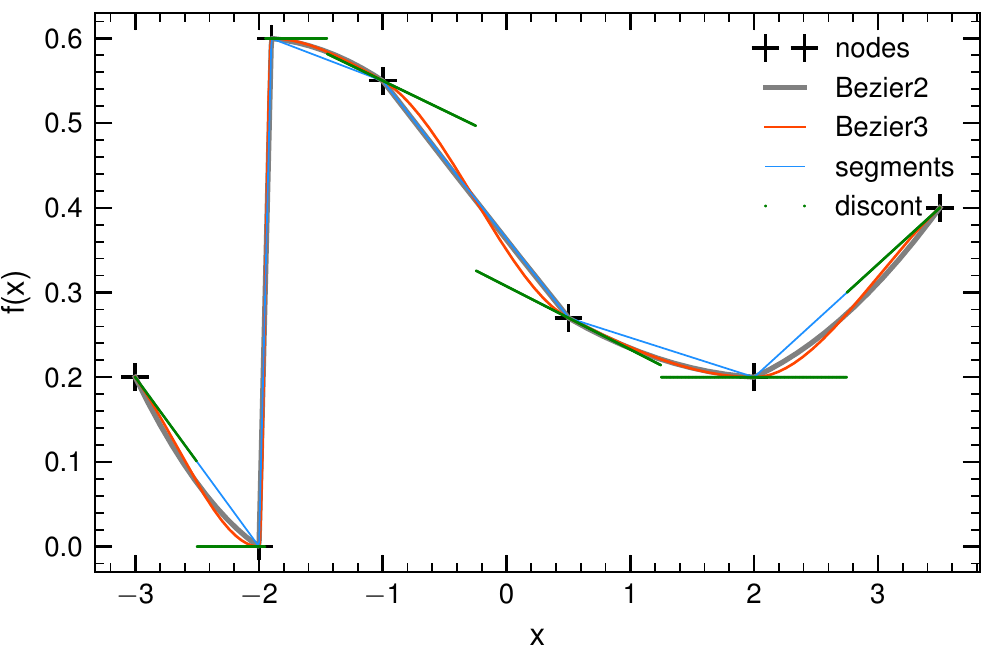}
\caption{Example of node interpolation using quadratic Bezier (solid gray), cubic Bezier (solid orange), straight segments (blue), and discontinous with slope delimiter (green dots). The node values are indicated with black crosses.}
\label{fig:interpolants}
\end{figure}

We allow the use of the following four types of nonovershooting interpolants:
\begin{enumerate}
\item Straight segments
\item Cuadratic Bezier splines\item Cubic Bezier splines
\item Discontinuous {grid-centered} interpolation with linear slope delimiter
\end{enumerate}
The exact implementation of these interpolants were described in detail in \citetads{2013ApJ...764...33D} and in \citetads{2016A&A...586A..42S}. Figure~\ref{fig:interpolants} illustrates an example showing fictitious node values and the interpolated curve using all these interpolants.
We allow the use of node parameterization in temperature, line-of-sight velocity, microturbulence, and the magnetic field vector ($B_\parallel$, $|B_\bot|$, $B_\chi$).

We also noticed that when integrating Eq.~(\ref{eq:hydro}), most inversion codes leave the boundary condition constant over the entire inversion. However, \citetads{1974SoPh...39...49S} and \citetads{2015ApJ...809L..30C} used observations in the Ca II H \& K lines and\ion{Mg}{ii}~h \& k lines, respectively, to derive ad hoc models that could produce similar spectra as in the observations. In both cases,  they needed between one and two orders of magnitude higher gas pressure in the upper chromosphere and transition region than that in the FALC quiet-Sun model (\citeads{1993ApJ...406..319F}). We investigated the response of the aforementioned lines to changes in the boundary condition for the hydrostatic equilibrium equation. 

\begin{figure*}
\centering
\includegraphics[width=\textwidth]{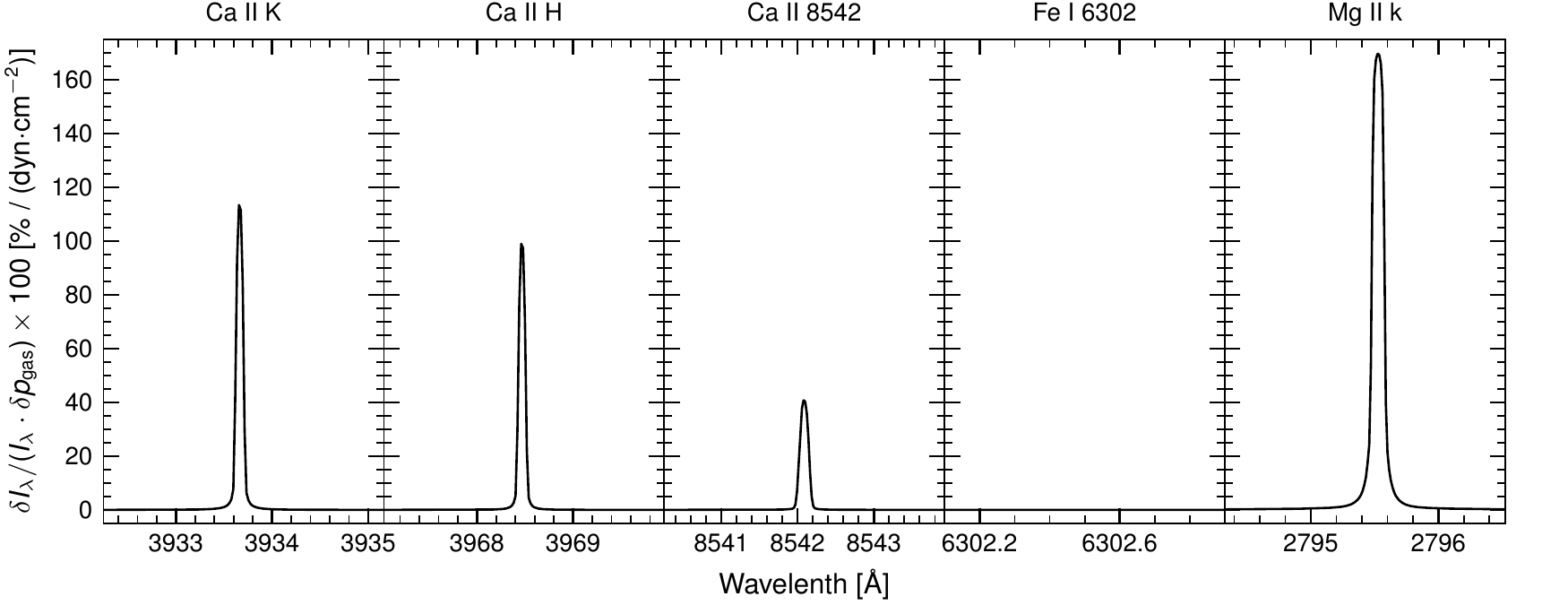}
\caption{Normalized response function to perturbations to gas pressure at the upper boundary condition of the atmosphere. }
\label{fig:pgasresp}
\end{figure*}

Fig.~\ref{fig:pgasresp} illustrates the response of \ion{Ca}{ii} H \&~K,  \ion{Ca}{ii}~854.2~nm, \ion{Fe}{i}~630.2~nm, and \ion{Mg}{ii}~k to perturbations in the gas pressure upper boundary. We used a plage model atmosphere that is very similar to that derived by \citetads{2015ApJ...809L..30C}. For this model, increasing the gas pressure at the boundary increases the line core intensity of all chromospheric lines but photospheric lines remain unaffected. Fig.~\ref{fig:pgasresp} also illustrates the differences in opacity among these lines. \ion{Mg}{ii}~k has more opacity than  \ion{Ca}{ii} H \&~K and all of these lines have more opacity than the \ion{Ca}{ii}~854.2~nm line.  

\begin{figure}
\centering
\includegraphics[width=\columnwidth]{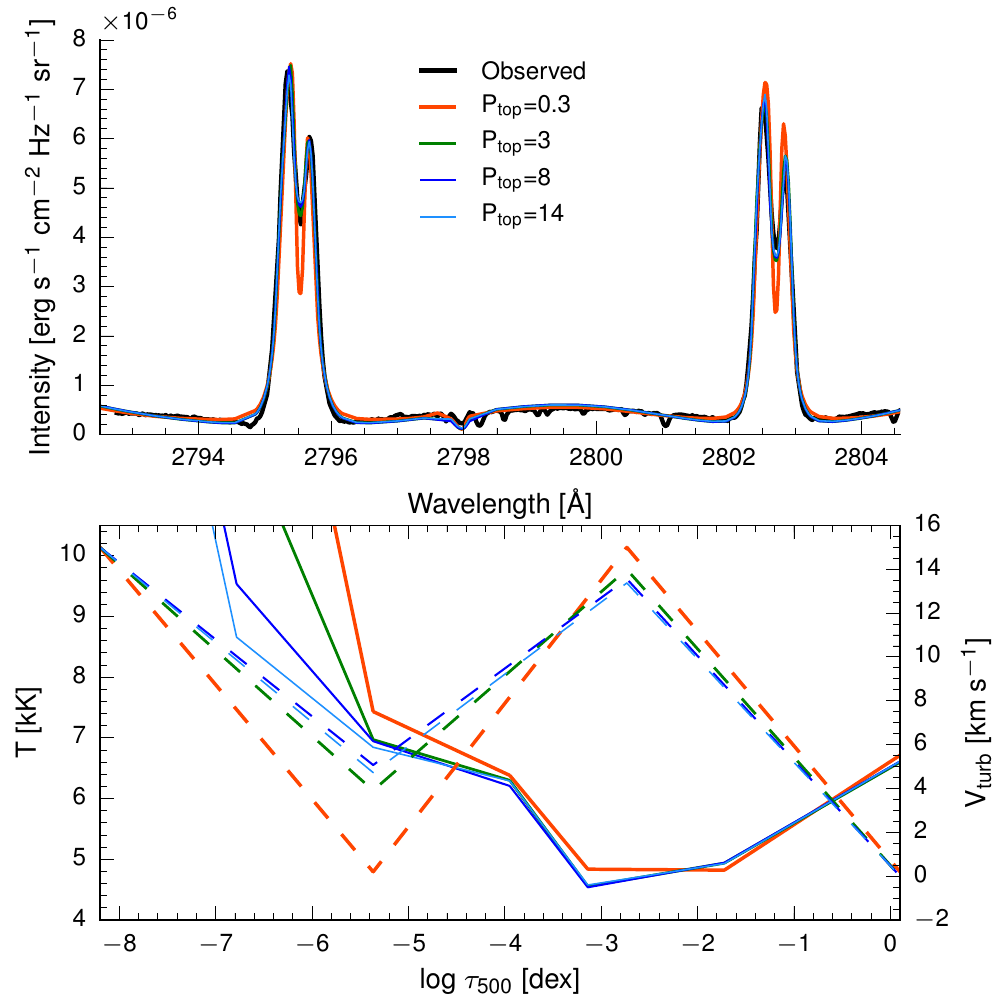}
\caption{Inversions of one IRIS plage observation performed with various gas pressure values at the upper boundary of the model atmosphere. Top: Fitted spectra (orange, green, navy and blue) and observed spectrum (black). {Bottom:} The reconstructed stratification of temperature and turbulent velocity for each inversion, represented using the same color coding as in the upper panel.}
\label{fig:pgas}
\end{figure}

Fig.~\ref{fig:pgas} illustrates the results of several inversions of a plage (IRIS) \ion{Mg}{ii}~h \&~k profile, using different values of the upper boundary gas pressure. For quiet-Sun values ($P_{top}=0.3$~dyn cm$^{-2}$, in red) the inversion cannot reproduce the enhanced line-core intensity of the observation. But when the boundary condition is increased to higher values ($P_{top} \sim 1.0$ dyn cm$^{-1}$), the line core intensity can be reproduced. By allowing the code to increase the gas pressure, the transition region can be moved to lower optical depth, where there is now more mass.

Fig.~\ref{fig:pgas} also illustrates that despite the degeneracy between the value of $P_{top}$ and the temperature gradient in the transition region, the inversion needs to have, at least, values that are one order of magnitude higher than in the quiet-Sun to reproduce the profile. We implemented the possibility of also adjusting the value of the upper boundary condition during the inversion as as free parameter. We found it more stable to implement it as a multiplicative factor to the upper boundary gas pressure. A good strategy to invert datasets that include quiet-Sun, sunspots and plage in the same field-of-view is to set the gas pressure to a value of approximately $P_{top} = 1.0$~dyn cm$^{-2}$ and let the code adjust the value to lower values if needed for the quiet-Sun areas. We discuss further how to do this and how this free parameter is regularized in \S\ref{sec:regul}.

\subsection{Parallelization scheme and I/O}
A C++ MPI-parallel code, STiC that follows a master-workers scheme. The parallelization is performed over pixels, assigning one worker to each vertical 1D model atmosphere. The master process performs I/O operations and distributes the workload among {worker} processes that are only used to process data. This scheme works particularly well when the time needed to process each package is not the same in all cases, allowing  balancing of the load of each worker on the fly. A similar scheme was used in NICOLE (\citeads{2015A&A...577A...7S}). We use the HDF5 library for data storage. This library is convenient because it allows storage of multiple named variables and metadata in one single file and it is supported by Python and IDL. 

\subsection{Regularizing Levenberg-Marquardt algorithm}\label{sec:LM}
The LM algorithm \citep[][]{10.2307/43633451, marquardt1963} facilitates a nonlinear least-squares fit of a model to observational data. The LM iteratively applies corrections to the parameters of a guessed model to minimize the difference between out synthetic and the measured data. This algorithm has been extensively used in solar inversion codes because it converges efficiently and it is particularly well suited for problems where the computation of derivatives is computationally expensive. 

In depth-stratified inversions finding the correct number of minimum number of free-parameters that allow the fitting of observations has been critical to avoid oscillatory behavior in the retrieved parameters. In this section we describe a regularizing LM algorithm that partly overcomes this issue. 

We begin by defining the {commonly used} merit function $\chi^2$ that is minimized, but in this case including a generic regularization term,
\begin{equation}
        \chi^2 (\bm{p}, \bm{x})= \frac{1}{N_{dat}} \sum_{k=1}^{N_{dat}} \bigg [\frac{o_k - s_k(\bm{p},x_k)}{\sigma_k} \bigg]^2 + \alpha r(\bm{p})^2, \label{eq:chi2gen}
\end{equation}
where $\bm{p}$ is a vector containing the $N_{par}$ parameters of the model, $s_k$ is the $k$-th prediction of our model (computed at the abscissa points $\bm{x}$), $o_k$ is the $k$-th measured data point and $\sigma_k$ is the error (or noise) of the $k$-th measurement, $\alpha$ is a weight for the regularization term, and $r(\bm{p})$ is a function that (in general) regularizes the problem by encouraging certain family of solutions that our algorithm prefers. In the following we work with normalized parameters, as these are numerically more stable. 

In our application we assume that we defined a number of individual penalty functions that can depend on different combinations of parameters contained in $\bm{p}$. That way the total penalty term is given by the sum of all ($N_{pen}$) individual penalties $r_n(\bm{p})$ as
\begin{equation}
        \chi^2 (\bm{p}, \bm{x})= \frac{1}{N_{dat}} \sum_{k=1}^{N_{dat}}\bigg[\frac{o_k - s_k(\bm{p}, x_k)}{\sigma_k}\bigg]^2 + 
        \sum_{n=1}^{N_{pen}} \alpha_n r_n(\bm{p})^2. \label{eq:chi2sum}
\end{equation}
The problem of Eq.~(\ref{eq:chi2sum}) is that now our figure to estimate the quality of the fits to the data also includes a term that depends on the model parameters themselves, while the standard definition of $\chi^2$ is normalized by the noise. Therefore it is particularly important to work with normalized model parameters within the LM part, which can be done by assuming a typical {norm} that scales the stratification of each physical parameter to values relatively close to unity {(e.g., $T=5000$~K, $v_{l.o.s}=6$~km~s$^{-1}$, $v_{turb}=6$~km~s$^{-1}$, $B_\parallel=1000$~G, $B_\bot=1000$~G, $B_\chi=\pi$ rad).}
The idea behind the LM algorithm is that, in each iteration $i$, we can find {corrections} ($\bm{\Delta p_i} = \bm{p} - \bm{p_i})$ to a set of model parameters ($\bm{p}$) that decrease our merit function $\chi^2$, so that $\chi^2(\bm{p_i}) > \chi^2(\bm{p_i} + \bm{\Delta p_i})$. 

At this point, one way to proceed would be linearizing Eq.~(\ref{eq:chi2sum}), but that would assume that all dependences with the model parameter are linear (see Appendix~\ref{ap:alternLM}). Instead, it is more appropriate to consider a second order Taylor expansion of the merit function around the current estimate of the parameters (e.g., \citeads{1992nrfa.book.....P}), i.e.,\begin{equation}
\chi(\bm{p}, \bm{x})^2 = \chi(\bm{p}_i, \bm{x})^2 + \bm{\Delta p}_i \nabla \bigg [ \chi(\bm{p}_i, \bm{x})^2\bigg ] + \frac{1}{2}\bm{\Delta p}_i \mathbf{D}\bm{\Delta p}_i, \label{eq:taylor}
\end{equation}
where $\mathbf{D}$ is the Hessian matrix of dimension $N_{par} \times N_{par}$. At the minimum, the derivative of Eq.~(\ref{eq:taylor}) must be zero, i.e.,
\begin{equation}
        \nabla\bigg [ \chi(\bm{p}, \bm{x})^2 \bigg ] =   \nabla \bigg [ \chi(\bm{p}_i, \bm{x})^2\bigg ] + \mathbf{D}\bm{\Delta p}_i = 0.\label{eq:dtaylor}
\end{equation}
The correction to our parameters at iteration $i$ is therefore given by the solution to the linear system of equations,
\begin{equation}
\mathbf{D}\bm{\Delta p}_i = - \nabla \bigg [ \chi(\bm{p}_i, \bm{x})^2\bigg ], \label{eq:sol}
\end{equation}
and the new estimate of the model parameters is therefore given by
\begin{equation*}
\bm{p}_{i+1} = \bm{p}_{i} + \bm{\Delta p}_i. 
\end{equation*}
If we now expand Eq.~(\ref{eq:sol}) using Eq.~(\ref{eq:chi2sum}), we only need to derive a formula for the Hessian and gradient of our merit function. Strictly speaking, the elements of the Hessian are given by
\begin{multline}
D_{jz} = \frac{\partial^2 \chi^2}{\partial p_z \partial p_j} =  \\ = 2\sum_k^{N_{dat}} \frac{1}{\sigma_k^2}\bigg [ \frac{\partial s_k}{\partial p_z}\frac{\partial s_k}{\partial p_j} - (o_k-s_k) \frac{\partial^2 s_k}{\partial p_z \partial p_j} \bigg ] +  2\sum_n^{N_{pen}} \alpha_n \frac{\partial^2 (r_n^2) }{\partial p_j \partial p_z}, \label{eq:Hess}
\end{multline}
but in the LM method it is customary to assume that the differences between the observed and synthetic data points $(o_k-s_k)$ are very small close to the minimum, and therefore to approximate that part of the Hessian with a linearized version that only depends on the first derivatives,
\begin{equation}
D_{jz} \approx 2\sum_k^{N_{dat}} \frac{1}{\sigma_k^2}\frac{\partial s_k}{\partial p_z}\frac{\partial s_k}{\partial p_j}  + 2\sum_n^{N_{pen}} \alpha_n \frac{\partial^2 (r_n^2) }{\partial p_j \partial p_z} = A_{jz}, \label{eq:Hess2}
\end{equation}
where we now denote the approximation to the Hessian matrix with $\mathbf{A}$, as is usually done in the literature. In general we should not assume a similar linearization of the penalty term. However, when the regularizing function $r_n$ has a linear dependence with the parameters, then such linearization is exact and we can compute that contribution to the Hessian using only the product of its Jacobian matrix terms (which we show in Appendix~\ref{ap:alternLM}). The latter is the reason why this algorithm can also be derived assuming a linear model of $\chi^2$ under this assumption. Since all the penalty functions that we consider in this paper fulfill this requirement, we continue using the linearized case.

The gradient of the merit function (the right hand term in Eq.~(\ref{eq:sol})) is trivially given by
\begin{equation}
 - \frac{\partial \chi^2}{\partial p_j} = 2\sum_k^{N_{dat}}\frac{1}{\sigma_k^2}\bigg [ (o_k-s_k) \frac{\partial s_k}{\partial p_j} \bigg ] - 2\sum_n^{N_{pen}} \alpha_n r_n \frac{\partial r_n}{\partial p_j}. \label{eq:grad}
\end{equation} 

Eq.~(\ref{eq:sol}) can be written in matrix form as
\begin{equation}
\mathbf{A} \bm{\Delta p}=  \mathbf{J}^T (\bm{o}-\bm{s}) - \mathbf{L}^T \bm{r}, \label{eq:mat}
\end{equation}
where $\mathbf{A}$ is the Hessian matrix, $\mathbf{J}$ is the Jacobian of the synthetic spectra, and $\mathbf{L}$ is the Jacobian of the regularization functions. We included the division by $\sigma_k$ in $\mathbf{J}$ and $(\bm{o}-\bm{s})$ and the a factor $\sqrt{\alpha_n}$ in the corresponding $\mathbf{L}$ and $\bm{r}$. The linearized approximate Hessian matrix, assuming a linear dependence of the penalty functions with the parameters, can be written as
\begin{equation}
\mathbf{A} =  \mathbf{J}^T  \mathbf{J} +  \mathbf{L}^T \mathbf{L}.
\end{equation}

In some situations, the linear system in Eq.~(\ref{eq:mat}) can lead to unstable solutions. Following Marquardt's insights, the diagonal of the Hessian matrix can be modified to stabilize the solution as
\begin{equation}
        \bar{A}_{ij} = 
        \begin{cases}
        (1+\lambda)A_{ii} & i=j,\\
        A_{ij} & i\neq j.
        \end{cases}
\end{equation}

The $\lambda$ parameter is a Lagrange multiplier that allows for switching between a steepest descent (when $\lambda$ is large) and a conjugate gradient method (when $\lambda$ is small). In our implementation we selected the value of $\lambda$ by doing a simple line search that brackets the optimal value of $\lambda$ and then we refined the optimal value with a parabola fit, which seems to work particularly well when regularization is included. 

So the final (linearized) equation that we need to solve is written as
\begin{equation}
 \bar{\mathbf{A}} \bm{\Delta p}=  \mathbf{J}^T (\bm{o}-\bm{s}) - \mathbf{L}^T \bm{r}. \label{eq:final}
\end{equation}
Comparing Eq.~(\ref{eq:final}) with a traditional LM implementation, there is a new term that modifies the Hessian matrix and an additional residual on the right-hand side. We are effectively changing the way the Hessian maps a given solution into the right-hand term. The extra residual term in the right-hand term balances the equality and provides insight into how the parameter corrections must be driven to minimize the penalty term as well.

This kind of $\ell-2$ regularization has been extensively used to solve ill-posed problems in stellar applications (e.g., \citeads{2002A&A...381..736P}). Other types of regularization have been included in the past in other inversion codes, and perhaps the closest implementation to our method can be found in NICOLE. However there are significant differences. Perhaps the main difference is that the penalty term in NICOLE is not squared in the definition of $\chi^2$, which changes completely the algebra of the problem from the beginning of the derivation, as the derivative of the penalty term respect to $\bm{ \Delta p}$ is different from ours after using the parabolic approximation to solve the problem in Eq.~(\ref{eq:taylor}). NICOLE must be applying $\ell-1$ regularization whereas our approach operates with $\ell-2$ norms. The regularization functions that we use are also rather different in nature than those in NICOLE (see \S\ref{sec:regul}).

To solve the linear system of equations in Eq.~(\ref{eq:final}), we use singular value decomposition (SVD). The {latter} is used in most inversion codes that are currently available to get the correction to the model parameters. In situations where $\mathbf{A}$ is rank deficient, SVD provides a least-squares fit to the solution of that system of equations. Additionally, very small singular values can be avoided. This way of solving the linear system is also a regularization method but, unlike using the penalty function, it operates on the projected total Hessian matrix, not on individual physical parameters. Therefore it is harder to understand how it affects individual physical parameters. 

\citetads{1992ApJ...398..375R} suggested checking the contribution of each singular value to the response of each physical parameter of the model and making sure that these contributions are filtered individually for each physical parameter. In our tests, the latter seems to mostly help with parameters that induce very small response in $\chi^2$ like the magnetic field azimuth, but proper weighting of the Stokes parameters can have a similar effect. Using SVD alone (without the other regularization terms) has the disadvantage that it does not particularly select any family of solutions, but instead filters the Hessian matrix in a way that removes unstable values in the system. In our experience both methods have slightly different effects and the combination of both methods greatly improves convergence.
Our calculations of SVD decomposition are performed using the excellent \CC\ Eigen-3 library \citep{eigenweb}. 

\subsubsection{Selection of the regularization weight $\alpha_n$}\label{sec:alpha}
The real challenge in this approach is how to chose the regularization weights $\alpha_n$ correctly. Too much weight affects the quality of the fits, as the problem is dominated by the penalty term. Too little regularization does not remove potential degeneracies in the solution. We surveyed the literature and there is a vast selection of different theoretical approaches to select the regularization weights, regardless of the fitting algorithm that is being employed \citep[see, e.g., ][]{kaltenbacher2008iterative, doicu2010numerical}. 

Perhaps one of the best (practical) insights into how to choose the regularization weight is provided by \citet{2017A&A...597A..58K}, based on the so-called L-curvature approach \citep{hansen1993}. The idea is to compare how the value of the square of the residual $||\bm{o}-\bm{s}||^2_2$ compares to the values of the penalty term $||r||^2_2$ for different values of $\alpha$ in a $\log - \log$ plot. Normally, for too small values of $\alpha$, the quality of the fits remains rather constant. Once the order of magnitude of the penalty term approaches the same order of magnitude as $||\bm{o}-\bm{s}||^2_2$, then the quality of the fits usually begins to degrade rapidly. The idea is to choose a value of $\alpha$ very close to that turnover point.

If we work with model parameters that are normalized to values around unity, for example by scaling the parameters by a norm, and if our estimate of the noise is adequate, then we should be able to choose $\alpha$ so that the penalty term remains slightly below unity. Now this is not exactly as {straightforward} as it sounds. {The temperature and magnetic field strength,} for example, {have} quite a large dynamic range compared to the other parameters. This is especially the case if our observations include lines that form under very different {physical conditions} because in that case, the model must accommodate all these {regimes} and the consequent gradients that connect them.


\citet{2017A&A...597A..58K} also indicates that some uncertainty in the choice of $\alpha$ cannot be avoided because we do not know the best fit a priori. However, we can perform test inversions on selected pixels to calibrate a good value. Normally we would need to be off by a factor $\times 10$ in order to see significant errors.
\begin{figure*}
\centering
\includegraphics[width=\textwidth]{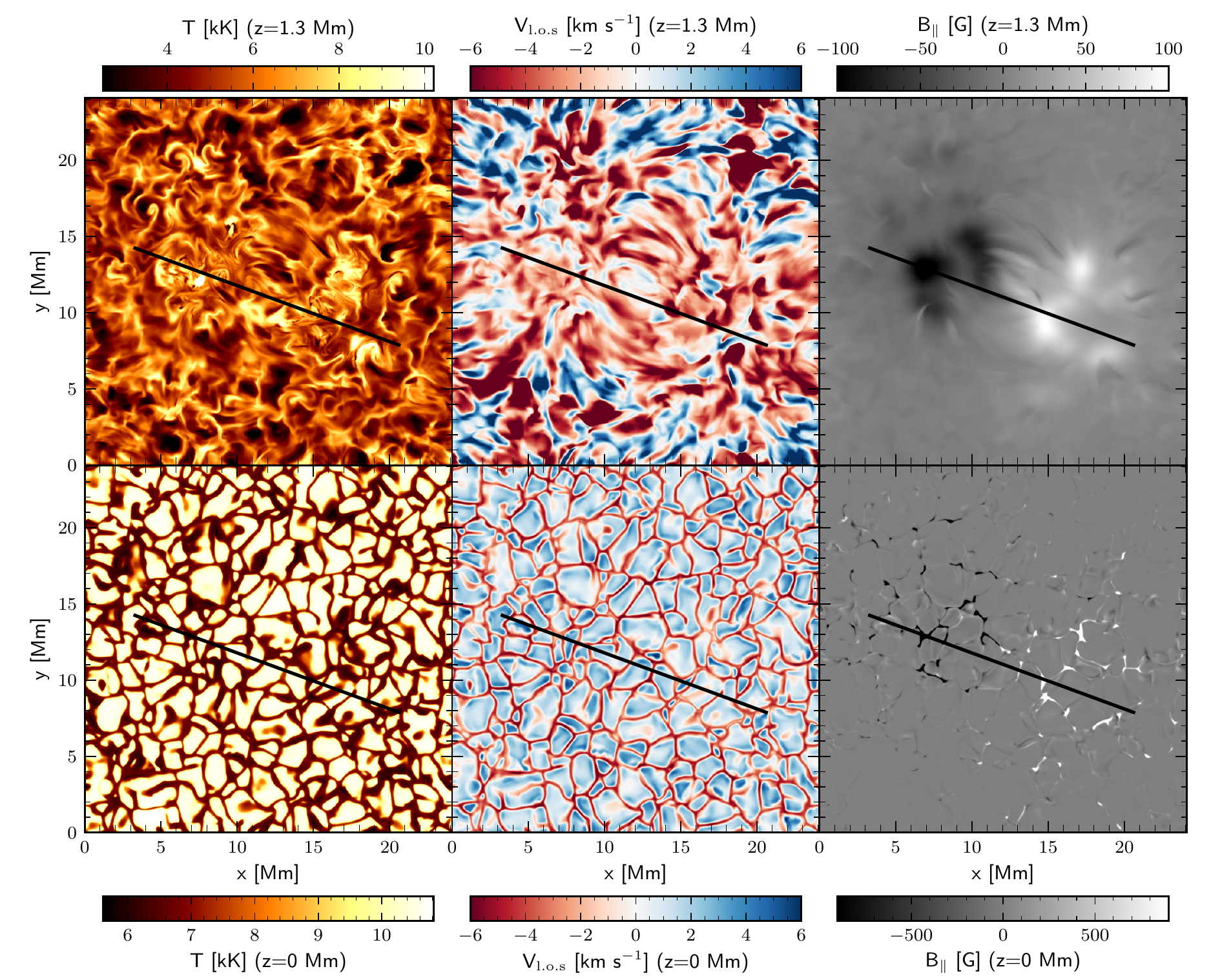}
\caption{Physical parameters from a snapshot of a 3D rMHD BIFROST simulation (\citeads{2016A&A...585A...4C}). From left to right, temperature, line-of-sight velocity and the vertical component of the magnetic field. The upper row shows a horizontal cut in the chromosphere at $z=1.3$~Mm and the lower row in the photosphere at $z=0$~Mm. The black line illustrates the region that we extracted to perform our inversion tests.}
\label{fig:bif2}
\end{figure*}

\begin{figure*}
\centering
\includegraphics[width=0.97\textwidth]{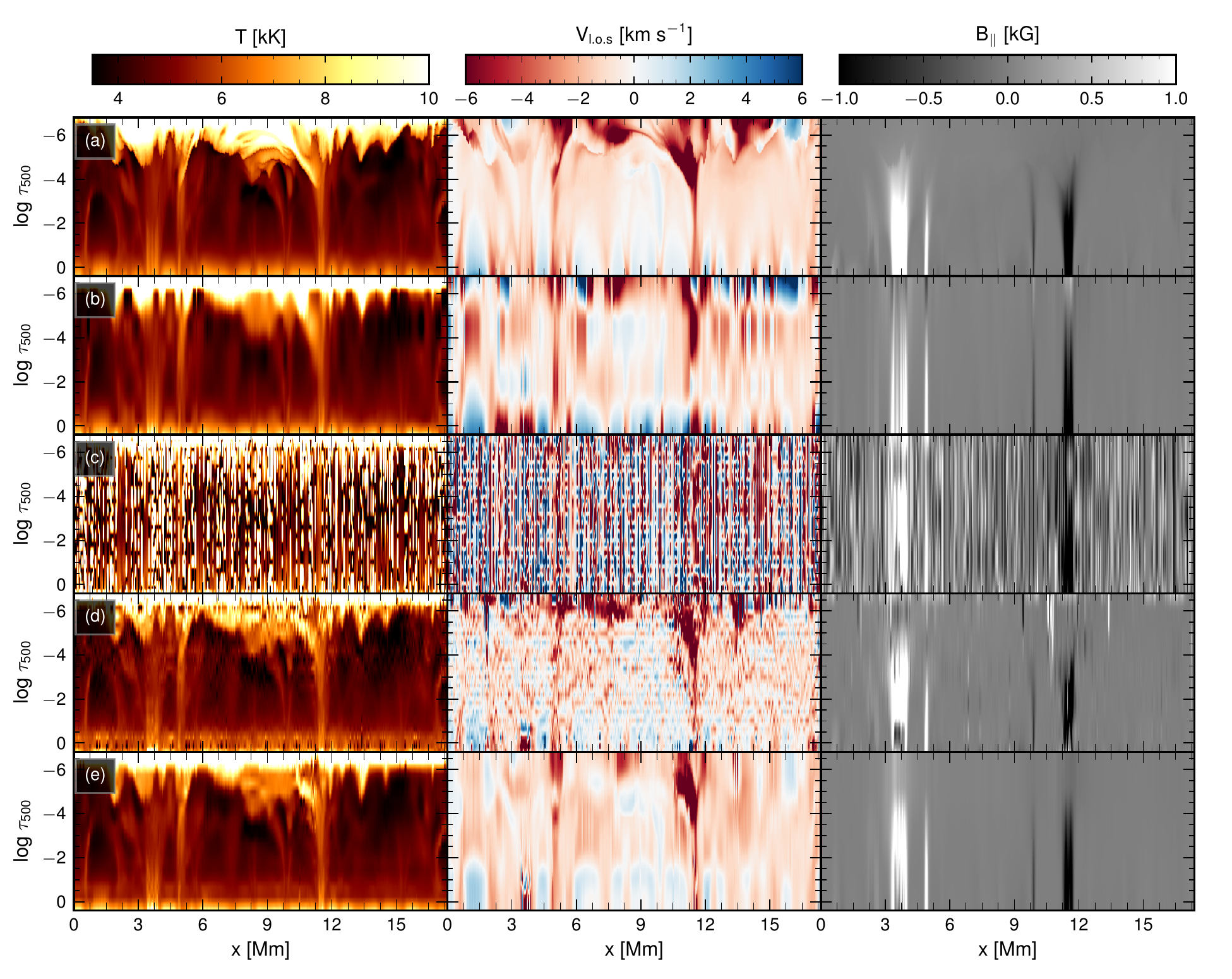}
\caption{Inversion of a vertical slice from a 3D rMHD simulation. The panels illustrate  from left to right the stratification in optical depth of temperature, line-of-sight velocity, and the vertical component of the magnetic field. {Row (a):} Original physical quantities from the MHD simulation. {Row (b):} Inversion computed without regularization and a small number of nodes. {Row (c):} Inversion computed without regularization and a very large number of nodes. {Row (d):} Inversion computed with understimated regularization and a very large number of nodes. {Row (e):} Inversion computed with a very large number of nodes and properly scaled regularization. The exact number of nodes of each experiment is indicated in Table~\ref{tab:nodes}.}
\label{fig:sim}
\includegraphics[width=0.97\textwidth]{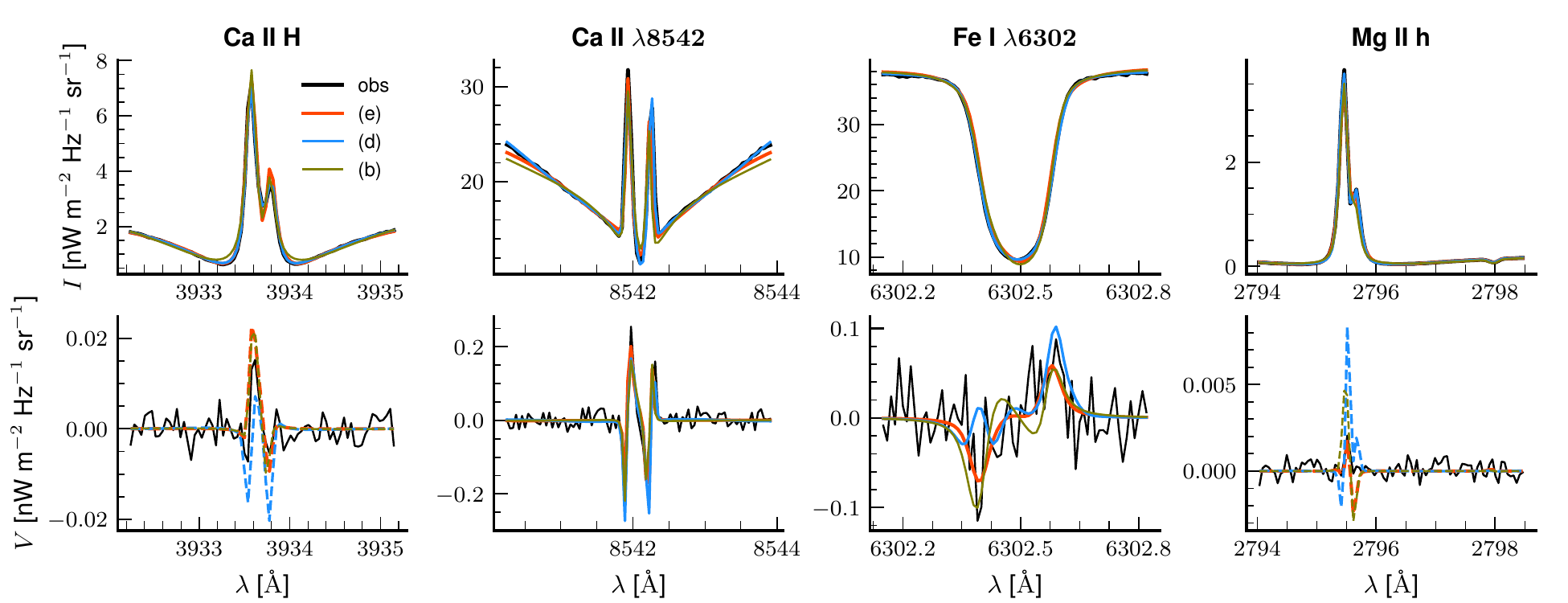}
\caption{Example fit of the different  inversions (row (e), $x=7.9$).  We only fitted Stokes $Q$, $U,$ and $V$ in the $\lambda 8542$, $\lambda 6301,$ and $6302$ lines. The fits for the \ion{Ca}{II}~H, \ion{Fe}{i}~$\lambda 6301,$ and \ion{Mg}{ii}~h lines are not shown because they are virtually identical to those in their partnering lines. At the considered noise level, all $Q$ and $U$ signals are below the noise level in all chromospheric lines.}
\label{fig:sim_fit}
\end{figure*}

\subsubsection{Regularization functions}\label{sec:regul}
If the inversion is performed with many degrees of freedom, the solution can become unstable, introducing oscillatory behavior in the derived parameters as a function of depth. Regularization techniques provide a natural way to discourage certain families of solutions by adding a penalty term to the definition of $\chi^2$, as generally expressed in Eq.~(\ref{eq:chi2gen}) with $r(\bm{p})$.

We implemented Tikhonov's regularization on the first derivative \citep{Tikhonov77}. Hereafter, the label $phyc$ indicates that a given vector or constant is related to one physical parameter (e.g., temperature). In this case for each interval between two consecutive nodes in one physical parameter ($k=1,...,N_{phyc}$, e.g., temperature) we define a regularization function (we note that the penalty function is not squared here, unlike in the definition of $\chi^2$) using the first derivative of the actual values of the physical parameter that we are considering ($\bm{p}_{phyc}$),
\begin{equation}
f_k(\bm{p}_{phyc}) = (p_{j}-p_{j-1}). \label{eq:tik}
\end{equation}
In this case, the derivatives of the regularization term relative to all parameters contained in $\bm{p}$ can be written as
\begin{equation}
\frac{\partial }{\partial p_i} f_k (\bm{p}_{phyc})= 
\begin{cases}
        1 & i=j,\\
        -1 & i=j-1,\\
        0 & \mathrm{otherwise},
\end{cases}
\label{eq:dtik}
\end{equation}
where $p_j$ are the values of the nodes for a given physical parameter. In this case, the block in ${H}$ corresponding to this physical variable only has nonzero terms in the diagonal and the band just below it. Alternatively, we could choose to penalize {changes} in the 
gradient of a variable, which is by definition what happens when we introduce wiggles in a curve. The latter can be attained by penalizing large values in the second derivative. For nonregular grids (nonequidistant node placement) we can define the penalty function as
\begin{equation}
                f_k=(\bm{p}_{phyc}) = (A \cdot p_{j+1} + B \cdot p_j + C \cdot p_{j-1}), \label{eq:dsec}
\end{equation}
where 
\begin{eqnarray*}
A &=& \frac{2}{\Delta x_{j+1}(\Delta x_{j} + \Delta x_{j+1})},\\
B &=& -\frac{2}{\Delta x_j(\Delta x_j \cdot \Delta x_{j+1})}, \\
C &=& \frac{2}{\Delta x_j(\Delta x_j + \Delta x_{j+1})},
\end{eqnarray*}
are expressed in terms of the node separation $\Delta x_{j+1} = x_{j+1} - x_j$ and $\Delta x_j = x_j - x_{j-1}$. The derivative of this penalty function becomes trivially
\begin{equation}
\frac{\partial }{\partial p_i} f_k(\bm{p}_{phyc}) =
\begin{cases}
A & i=j+1,\\
B & i = j,\\
C & i= j-1,\\
0 & \mathrm{otherwise}.
\end{cases}
\label{eq:dconst}
\end{equation}

Another useful form of regularization is to penalize deviations of the stratification of a parameter from a constant value $v$,
\begin{equation}
f_k(\bm{p}_{phyc}) = (p_j - v),
\label{eq:const}
\end{equation}
where $v$ is a constant {expected value}. The derivative, which  in this case is trivial, is written as
\begin{equation}
\frac{\partial }{\partial p_i} f_k(\bm{p}_{phyc}) = p_j 
\begin{cases}
1 & i=j,\\
0 & i\neq j.\\
\end{cases}
\label{eq:dconst2}
\end{equation}
If $v$ is taken to be mean of all elements in $\mathbf{p}_{phyc}$ (denoted as $\bar{p} = \sum_{j=1}^{N_{phyc}}\, p_j$), then the derivative must also account
for the fact that changing one value $p_j$ also changes the value of $\bar{p}$, by including the derivative of $\bar{p}$ written as
\begin{equation}
\frac{\partial }{\partial p_i} f_k(\bm{p}_{phyc}) = 
\begin{cases}
1-\frac{1}{N_{phyc}} & i=j,\\
-\frac{1}{N_{phyc}} & i\neq j.
\end{cases}
\label{eq:dmean}
\end{equation}
Eq.~(\ref{eq:dconst}) obviously yields a diagonal block, whereas in Eq.~(\ref{eq:dmean}) all block elements are nonzero because all nodes contribute to the mean value. We note that Eq.~(\ref{eq:dtik}), (\ref{eq:dconst}), (\ref{eq:dconst2}), and (\ref{eq:dmean}) do not contain the constant factor $\alpha$ that multiplies each of the penalty functions.

Eq.~(\ref{eq:dtik}) and (\ref{eq:dsec}) are both capable of removing spurious oscillatory behavior and {wiggles} from the stratification of a parameter, but they operate in different ways. The former prefers to have small gradients in the solution whereas the latter only penalizes changes in the gradient itself. Similarly, the main difference between Eq.~(\ref{eq:tik}) and Eq.~(\ref{eq:const}) is that the former encourages smoothly varying solutions as a function of depth, whereas the latter does not necessarily encourage smoothness but discourages large deviations from the selected constant value. Eq.~(\ref{eq:dsec}) allows for a larger dynamic range in the stratification of a variable than Eq.~(\ref{eq:dtik}) and it is particularly well suited for variables such as temperature, where we may have to include the transition region. 

If the user decides to allow for adjustments in the gas pressure at the upper boundary ($P_{top}$,) we allow for regularizing the multiplicative factor of $P_{top}$ with penalties to deviations from a value of 1. That way, if the input model assumes an upper chromosphere value of $P_{top}~1.0$ dyn cm$^{-2}$, the code only increases or decrease this value when it actually improves the value of $\chi^2$, but {the algorithm will try} to adjust the temperature gradient if possible. We are basically selecting to fit as much as we can with changes to the temperature stratification. 

\subsubsection{Numerical experiment}\label{sec:regnumn}
We performed a numerical test with a vertical slice extracted from snapshot 385 from a public 3D rMHD simulation (\citeads{2016A&A...585A...4C}). Snapshots from this simulation have been used extensively in recent years (\citeads{2013ApJ...772...89L}; \citeads{2013ApJ...778..143P}; \citeads{2013ApJ...764L..11D}; \citeads{2014ApJ...784L..17L}; \citeads{2015ApJ...803...65S}; \citeads{2015ApJ...814...70R}) as it was made publicly available as part of the NASA IRIS mission (\citeads{2014SoPh..289.2733D}). This simulation includes the solar photosphere, chromosphere, and corona, allowing us to prepare a meaningful test case for the code. {In order to speed up our calculations, we selected a vertical cut of the snapshot, indicated in Fig.~\ref{fig:bif2} with a black line over the field of view. This line connects two patches of opposite polarity and it is aligned with fibril-like structures that are visible in the magnetic field and line-of-sight velocity panels. In pixel coordinates, the slit extends from $(x_0,y_0) = (69,298)$ to $(x_1,y_1)=(431,165)$.}

\begin{table}
\caption{Synthetic observations from a 3D rMHD simulation. The profiles have been convolved with a {Gaussian} PSF of FWHM of twice the sampling (FWHM$=2 \times \delta\lambda$). We note that the actual synthesis is {performed using} a finer wavelength grid in order to have accurate convolutions with the instrumental profile, but the extra points do not contribute to the inversion.}             
\label{tab:syn}      
\centering
\begin{tabular}{l |c c c}          
\hline\hline                        
Line & $\lambda_0$ & $\delta \lambda$ [m\AA]&  $\Delta \lambda$ [\AA\ from $\lambda_0$] \\
\hline                                   
\ion{Mg}{ii}~k & 2795.528 & 50.0 & $(-1.5, +2.86)$ \\
\ion{Mg}{ii}~h & 2802.705 & 50.0 & $\pm 0.93$\\
\ion{Ca}{ii}~K & 3933.664 & 39.5 & $\pm 1.4$\\
\ion{Ca}{ii}~H & 3968.469 & 39.5 & $\pm 1.4$\\
\ion{Fe}{i}~$\lambda 6301$ & 6301.501  & 25.0 & $\pm 0.27$\\
\ion{Fe}{i}~$\lambda 6302$ & 6302.493  & 25.0 & $\pm 0.27$\\
\ion{Ca}{ii}~$\lambda 8542$ & 8542.091 & 50.0 & $\pm 1.8$\\
\hline                                             
\end{tabular}
\end{table}

We synthesized spectra in the \ion{Mg}{ii}~h\&k, \ion{Ca}{ii}~H\&K, the \ion{Ca}{ii}~8542~\AA \ line and \ion{Fe}{i}~6301 \& 6302 \AA\ lines. This setup is representative of a co-observation between IRIS and the CRISP and CHROMIS instruments at the Swedish 1 m Solar Telescope (\citeads{2006A&A...447.1111S}; \citeads{2008ApJ...689L..69S}), which have been rather common since the launch of IRIS in 2014. Table \ref{tab:syn} summarizes the wavelength coverage of each line and the assumed spectral resolution.

In this test we want to show the usefulness  of regularization, especially to dig more detail out of the model atmosphere. In this case the magnetic field information is retrieved only from the $\lambda 8542$, $\lambda 6301$, and $\lambda 6302$ lines. In inversion runs, the depth resolution is set by the number of nodes that are being employed in a given physical parameter. The question however is how many nodes can we actually constrain during the inversion. Part of the answer is provided by the exact number of spectral lines that we observed, their sensitivities to different parts of the atmosphere and the spectral resolution of the observations. When the number of nodes is overestimated, the solution of the inversion shows oscillatory behavior and in extreme cases, the problem fails to converge at all. That is the reason why we normally must find {a setup that allows for reproducing the observed spectra with the lowest number of degrees of freedom} (\citeads{2017SSRv..210..109D}). 

An example of this effect is shown in Fig.~\ref{fig:sim}, where we illustrated a number of inversions experiments that are summarized in Table~\ref{tab:nodes}. {These results correspond to a single cycle inversion initialized with the same model atmosphere in all columns}. Rows (b) and (c) represent an inversion with a very limited number of nodes (upper-middle) and a case where the code failed completely to converge when we used an unrealistically large number of nodes (row (c)). In the former case we used 7 nodes in temperature and 4 nodes in line-of-sight velocity, whereas in the latter case we used 22 nodes in temperature and line-of-sight velocity. 

\begin{table}
\caption{Number of nodes used in our inversion setups. The results are shown in Figure~\ref{fig:sim}.}              
\label{tab:nodes}      
\centering
\begin{tabular}{c |c c c c c c | c}          
\hline\hline                        
Experiment &  $T$ &  $v_{l.o.s}$ & $v_{turb}$ & $B_\parallel$ & $|B_\bot|$ & $B_\chi$ & $\alpha$ \\
\hline                                   
(b) & 7 & 4 & 0 & 3 & 2 & 1 & 0 \\
(c) & 22 & 22 & 0 & 5 & 3 & 2 & 0\\
(d) & 22 & 22 & 0 & 5 & 3 & 2 & 0.1\\
(e) & 22 & 22 & 0 & 5 & 3 & 2 & 100\\
\hline                                             
\end{tabular}
\end{table}

Adding regularization helps the convergence rate significantly and it removes the oscillations significantly. Row (d) illustrates what happens when we add regularization but the regularization term is heavily understimated. The code manages to converge to a solution that resembles the original model, but wiggles are present all over the three physical quantities. When the right amount of regularization is added (see \S\ref{sec:alpha}), the problem converges to a solution that resembles the original model, as shown in row (e). In this case we applied penalty terms to the second derivative of the stratification of temperature as well as to the first derivative of the line-of-sight velocity and magnetic field. In principle there is not much difference among these two types, except that when we applied the latter to the temperature stratification, the temperature of the transition region was lower because it did not have a sufficiently relevant impact in $\chi^2$ but it lowered the penalty term in that case. If we had included transition region lines, we think this effect would probably not be there, although we are not probing this point here. 

Fig.~\ref{fig:sim_fit} shows the fits at $x=7.9$, where the line profiles have similar shapes to those observed. The fits are good in most lines and in all cases, except in Stokes~$V$ where \ion{Mg}{ii}~k and \ion{Ca}{ii}~K were not included in full-Stokes mode in the inversion. All fits capture the global shape of the line, but the details are better fitted in the regularized cases. The differences between the blue and red curves and are harder to judge by looking at individual pixels, but the values of $\chi^2$ are statistically very similar.

The case that we tested in this work has perhaps way too many degrees of freedom for being a realistic case, but it serves to show the power of regularization. It also illustrates that the method does not always converge entirely. For example, in the photosphere at $x\approx 3.5$~Mm the model has an artifact. At that location the \ion{Fe}{i} lines are split due to the Zeeman effect and in this case the algorithm gets stuck in a local minimum where the magnetic field is small in the photosphere and the line is broadened by having wiggles in velocity. In our experience, these kind of artifacts usually appear when the initial guessed model is quite far from the real solution and there are a very large number of nodes . A good strategy to avoid these artifacts is to perform a first cycle with less nodes and then restart the inversion from the solution of the latter, but with more degrees of freedom. The latter approach was already introduced by \citetads{1992ApJ...398..375R} and it speeds up the whole inversion process considerably as the response functions of that first cycle are faster to compute because the inversion is re-initialized from a closer solution to the minimum.

\section{Conclusions}\label{sec:conc}
We have developed a new inversion code that builds upon the ideas used in the SIR and NICOLE inversions codes. 
For the first time STiC allows us to consider lines from 
different atomic species while including PRD effects. The latter development allows the inclusion of
lines that sample the upper chromosphere and that also set stronger physical constraints in the mid and lower chromosphere.

We implemented $\ell-2$ regularization in our LM algorithm. The latter is introduced in the approximation to the
Hessian matrix directly, whereas other regularization techniques can be applied directly by projecting the parameters with a regularization operator (e.g., \citeads{2015A&A...577A.140A}; \citeads{2016A&A...590A..87A}). In this paper we show that regularization helps dig more detail out of the inversion by allowing the inclusion of more degrees of freedom, while getting rid of erratic oscillatory behavior. It also improves the convergence rate of the algorithm, even when the regularization amount is underestimated. The gain is particularly large in problems with particularly large number of nodes and atoms.

A word of caution seems appropriate though, as inversions codes always provide a result. It is up to the user not to over interpret those results and to check the robustness of the inversions. A good way to do so is to have a clear idea of what aspect of our problem we want to solve with inversions.

In our opinion, future developments should focus on the elegant solutions shown by \citetads{2017A&A...601A.100M}, where the response functions are computed analytically instead of by finite differences, although at the moment their formalism is not mature enough to include PRD lines.

STiC is publicly available to the community at \url{https://github.com/jaimedelacruz/stic}.

\begin{appendix}
\onecolumn

\section{Approximation of the cubic Bezier interpolation coefficients with Pad\'e interpolants}\label{sec:pade}
{ Most high order integration schemes can suffer from numerical precision issues for very small optical-depth values, even when the computations are performed in double precision. Therefore, most implementations switch to a third order Taylor expansion of the integration coefficients and of the exponential when $\tau_{uc} < 0.05$ (e.g., \citeads{2013A&A...549A.126I}).}

{We tested a similar approach with Pad\'e approximants and the resulting curve preserves five-digit accuracy up to $\tau_{uc} = 0.8$ in all integration coefficients and in the exponential. The third order Taylor expansion can only keep the same accuracy in all parameters up to $\tau_{uc}=0.046$ in our tests. The latter could be used to restrict even more the range in which the (expensive) exponential term needs to be computed.} For $\tau_{uc}<0.8$, the Pad\'e approximation of the cubic Bezier coefficients are written as
\begin{eqnarray*}
  \mathrm{e}^{-\tau_{uc}} &=& \frac{1-0.5\tau_{uc} + 0.1\tau_{uc}^2 - 8.33333\times10^{-3}\tau_{uc}^3}{1+0.5\tau_{uc} + 0.1\tau_{uc}^2 + 8.33333\times10^{-3}\tau_{uc}^3}.\nonumber\\
  \alpha &=& \frac{\frac{1}{4}\tau_{uc} - \frac{1873}{27030}\tau_{uc}^2 + \frac{9667}{1441600}\tau_{uc}^3}{1+\frac{7066}{13515}\tau_{uc} + \frac{1611}{14416}\tau_{uc}^2 + \frac{2353}{227052}\tau_{uc}^3},\nonumber\\
\beta  &=& \frac{\frac{1}{4}\tau_{uc} + \frac{1}{30}\tau_{uc}^2 + \frac{1}{480}\tau_{uc}^3}{1 + \frac{1}{3}\tau_{uc} + \frac{1}{24}\tau_{uc}^2+\frac{1}{504}\tau_{uc}^3},\\
\gamma &=& \frac{\frac{1}{4}\tau_{uc} -  \frac{89}{2360}\tau_{uc}^2 +\frac{619}{226560}\tau_{uc}^3 }{1 + \frac{53}{118}\tau_{uc} +  \frac{911}{11328}\tau_{uc}^2 + \frac{479}{79296}\tau_{uc}^3},\nonumber\\
\varphi &=& \frac{\frac{1}{4}\tau_{uc} -  \frac{1}{310}\tau_{uc}^2 +\frac{11}{14880}\tau_{uc}^3 }{1 + \frac{12}{31}\tau_{uc} +  \frac{43}{744}\tau_{uc}^2 + \frac{3}{868}\tau_{uc}^3}\nonumber.
\end{eqnarray*}

It is not always clear that these equations are faster to compute than the actual interpolation, but at least the approximation of the exponential can be combined with the real coefficients for larger values of $\tau_{uc}$.

\section{An alternative derivation of the regularizing LM algorithm}\label{ap:alternLM}
An alternative way to derive the regularizing LM algorithm is to linearize the dependence of Eq.~(\ref{eq:chi2sum}) respect to the model parameters.
In order to find {corrections} ($\bm{\Delta p})$ to a set of model parameters ($\bm{p}$) that decrease our merit function $\chi^2$, so that $\chi^2(\bm{p}) > \chi^2(\bm{p} + \bm{\Delta p})$, we perform a linear Taylor expansion of the merit function around the current value of the parameters. Given that we are assuming a nonlinear case and therefore we need to iterate the solution of our set of parameters, we can linearize the expression for $\chi^2(\bm{p} + \bm{\Delta p})$, assuming that $\bm{\Delta p}$ is sufficiently small in each iteration, i.e.,\begin{eqnarray}
        s_i(\bm{p}+\bm{\Delta p}) &=& s_i(\bm{p}) + \bm{j}_i^T\bm{\Delta p},\label{eq:linS}\\
        r_j(\bm{p}+\bm{\Delta p}) &=& r_j(\bm{p}) + \bm{h}_j^T\bm{\Delta p}\label{eq:linR},
\end{eqnarray}
where $\bm{j}_i$ is the Jacobian (vector of $N_{par}$ elements) of the synthetic spectrum $s_i(\bm{p}, x_i)$  and ${h}_j$ is the Jacobian (vector of $N_{par}$ elements) of a single $r_j(\bm{p})$.
In principle, we can replace Eq.~(\ref{eq:linS}) and (\ref{eq:linR}) into Eq.~(\ref{eq:chi2sum})as follows:
\begin{equation}
        \chi^2 (\bm{p}+\bm{\Delta p}, \bm{x})   = \frac{1}{N_{dat}} \sum_{i=1}^{N_{dat}} \bigg[\frac{o_i - s_i(\bm{p},x_i) -\bm{j}_i^T\bm{\Delta p}}{\sigma_i}\bigg]^2 
        + \sum_{j=1}^{N_{pen}}\alpha_j \bigg[  r_j(\bm{p})+ \bm{h}^T\bm{\Delta p}\bigg]^2. \label{eq:chi2}
\end{equation}
We can rewrite Eq.~(\ref{eq:chi2}) in matrix form, which simplifies enormously the algebraic manipulations and the notation. In that case, we define ${J}$ as the full Jacobian matrix for all data points with dimensions $(N_{par}, N_{dat})$ and ${H}$ is a matrix with dimensions $(N_{par},N_{pen})$. Each column of ${H}$ contains the derivatives of one individual penalty function relative to all parameters in $\bm{p}$. The penalty functions are contained in the components of vector $\bm{r} = (r_1, r_2, ..., r_{N_{pen}})$; 
\begin{equation}
 \chi^2 = \bigg [\bm{o} - \bm{s} - {J}^T\bm{\Delta p}\bigg ]^2 + \bigg [ \bm{r} + {H}^T\bm{\Delta p}\bigg ]^2 = \bigg [\bm{o} - \bm{s} - {J}^T\bm{\Delta p}\bigg ]^T  \bigg[\bm{o} - \bm{s} - {J}^T\bm{\Delta p}\bigg ]+ \bigg [ \bm{r} + {H}^T\bm{\Delta p}\bigg ]^T \bigg [ \bm{r} + {H}^T\bm{\Delta p}\bigg ]. \label{eq:chimat}
\end{equation}
We implicitly hid the division by the noise in all relevant matrices, and we included the $\alpha$ factors in the vector $\bm{r}$. If we equal to zero the derivative of Eq.~(\ref{eq:chimat}) with respect to $\bm{\Delta p}$ and after performing some basic matrix algebra, we can find the corrections $\bm{\Delta p}$ that minimize our merit function.
 
 Defining the modified approximate Hessian matrix,
\begin{equation}
        \mathbf{A} = \mathbf{J} \cdot \mathbf{J}^T + \mathbf{H} \cdot \mathbf{H}^T,
\end{equation}
then the corrections to our current estimate of the parameter are given by the following linear system of equations; {these corrections} include the effect of our regularizing functions $r(\bm{p})$ and their derivatives as follows:
\begin{equation}
        \mathbf{A} \cdot\bm{ \Delta p} = \mathbf{J} \cdot (\bm{o} - \bm{s}) - \bigg [\mathbf{H} \cdot \mathbf{r}\bigg ].\label{eq:LMregNo}
\end{equation}
Eq.~(\ref{eq:LMregNo}) is very similar to the linear system usually considered in a standard LM algorithm. We simply modified the Hessian matrix and added an extra term to the {residue} on the right-hand side to account for the regularization terms. By definition, the linear approximation of the Hessian is automatically recovered, and the only limitation compared to the full algorithm in \S\ref{sec:LM} is that the penalty terms must have a linear dependence with the model parameters.
\end{appendix}
\twocolumn

\begin{acknowledgements}
We are most thankful to N. Piskunov for sharing his LTE EOS with us. We thank J. Pires Bj\o rgen for his valuable help with the implementation of the non-LTE hydrogen ionization routines. We greatly appreciate the comments provided by I. Milic (the referee) who helped to improve the quality of this paper.
JdlCR is supported by grants from the Swedish Research Council (2015-03994), the Swedish National Space Board (128/15). This project has received funding from the European Research Council (ERC) under the European Union's Horizon 2020 research and innovation programme (SUNMAG, grant agreement 759548). SD and JdlCR are supported by a grant from the Swedish Civil Contingencies Agency (MSB).
This research was supported by the CHROMATIC grant of the Knut and Alice Wallenberg foundation.
 Computations were performed on resources provided by the Swedish National
 Infrastructure for Computing (SNIC) at the PDC Centre for High Performance Computing (PDC-HPC)
 at the Royal Institute of Technology in Stockholm as well as recourses at the High Performance Computing Center North (HPC2N).
This study has been discussed
within the activities of team 399 'Studying magnetic-field-regulated heating in the solar chromosphere' at the International Space Science Institute (ISSI) in Switzerland.
The Swedish 1-m Solar Telescope is operated on the island of La Palma by the Institute for Solar Physics of Stockholm University in the Spanish Observatorio del Roque de los Muchachos of the Instituto de Astrof\'isica de Canarias.
The Institute for Solar Physics is supported by a grant for research infrastructures of national importance from the Swedish Research Council (registration number 2017-00625).
 \end{acknowledgements}

\bibliographystyle{aa} 
\bibliography{references}

\end{document}